\documentclass[aps,prl,twocolumn,showpacs,groupedaddress]{revtex4-1}

\usepackage{graphicx}
\usepackage{amsmath}
\usepackage{amssymb}

\usepackage{color}

\begin{document}

\title{Magnetotransport signatures of the proximity exchange and spin-orbit couplings in graphene}
\author{Jeongsu Lee}
\email[]{jeongsu.lee@physik.uni-regensburg.de}
\author{Jaroslav Fabian}
\affiliation{Institute for Theoretical Physics, University of Regensburg, 
93040 Regensburg, Germany}

\begin{abstract}
Graphene on an insulating ferromagnetic substrate---ferromagnetic insulator
or ferromagnetic  metal with a tunnel barrier---is expected to exhibit giant proximity
exchange and spin-orbit couplings. We use a realistic transport model
of charge-disorder scattering and solve the linearized Boltzmann equation 
numerically exactly for the anisotropic Fermi contours of modified Dirac electrons to find magnetotransport signatures of these proximity effects: proximity anisotropic magnetoresistance, inverse spin-galvanic effect, and the planar Hall resistivity. We establish the corresponding anisotropies due to the exchange
and spin-orbit coupling, with respect to the magnetization orientation. We also present 
parameter maps guiding towards optimal regimes for observing transport magnetoanisotropies
in proximity graphene.
\end{abstract}

\pacs{
72.20.Dp	
72.80.Vp	
73.22.Pr	
73.63.-b	
}

\maketitle

Dirac electrons in pristine graphene have wesizableak spin-orbit coupling 
\cite{Gmitra:2009fh} and no magnetic moments, limiting prospects for spintronics \cite{Zutic:2004fo}.  This can be partly
remedied by functionalizing graphene with adatoms and admolecules, which can induce
sizable {\it local} magnetic moments and spin-orbit coupling, leading to marked 
spin transport fingerprints \cite{Kochan:2014hz, Bundesmann:2015bh, VanTuan:2016ct, Ferreira:2014bc,
Wilhelm:2015cl, Thomsen:2015jr}. A more systematic and, important, spatially {\it uniform} way to induce spin properties in graphene is by proximity effects. Being essentially a surface (or two), graphene can ``borrow'' 
properties from its substrates. Placing graphene on a slab of a
ferromagnetic insulator, or a ferromagnetic metal with a tunnel barrier, is expected to 
induce giant proximity exchange as well as spin-orbit coupling in the Dirac electron band
structure. This is supported by first principles calculations  \cite{Yang:2013jd, Qiao:2014fw, Lazic:2016dh, Crook:2015ea, Zollner:2016wq}   as well as by recent 
experiments on graphene on yttrium iron garnet \cite{Wang:2015gd, Leutenantsmeyer:2016vh} and on graphene on EuS \cite{Wei:2016hw}. In effect, proximity graphene on ferromagnetic substrates should be an ultimately thin ferromagnetic layer, with giant spin-orbit coupling, forming a perfect 
playground for both spintronics experiment and theory  \cite{Han:2014dc}. 

An important question is: what transport ramifications can we expect in such a magnetic
graphene with strong spin-orbit coupling? On one hand, in ferromagnetic metals the exchange coupling is typically much greater than spin-orbit coupling. On the other hand, 
in semiconductor heterostructures, 
which are the best case studies for structure-induced spin-orbit coupling in its
transport signatures \cite{Trushin:2007kg, Chalaev:2008gk}, there is no 
ferromagnetic exchange and spin splitting can be due to the Zeeman interaction which is, 
for realistic values of magnetic field, 
much weaker than spin-orbit coupling. Proximity graphene should be intermediate between those two extremes: the proximity exchange and spin-orbit couplings are expected  to be similar, on the order of 1 - 10 meV
\cite{Avsar:2014ex, Yang:2013jd, Qiao:2014fw}.  Perhaps the main 
effect of the interplay of exchange and spin-orbit couplings---magnetotransport anisotropies---should 
be well pronounced and make for useful, experimentally testable signatures of the spin proximity effects.

\begin{figure}[b]
\includegraphics[width=\columnwidth]{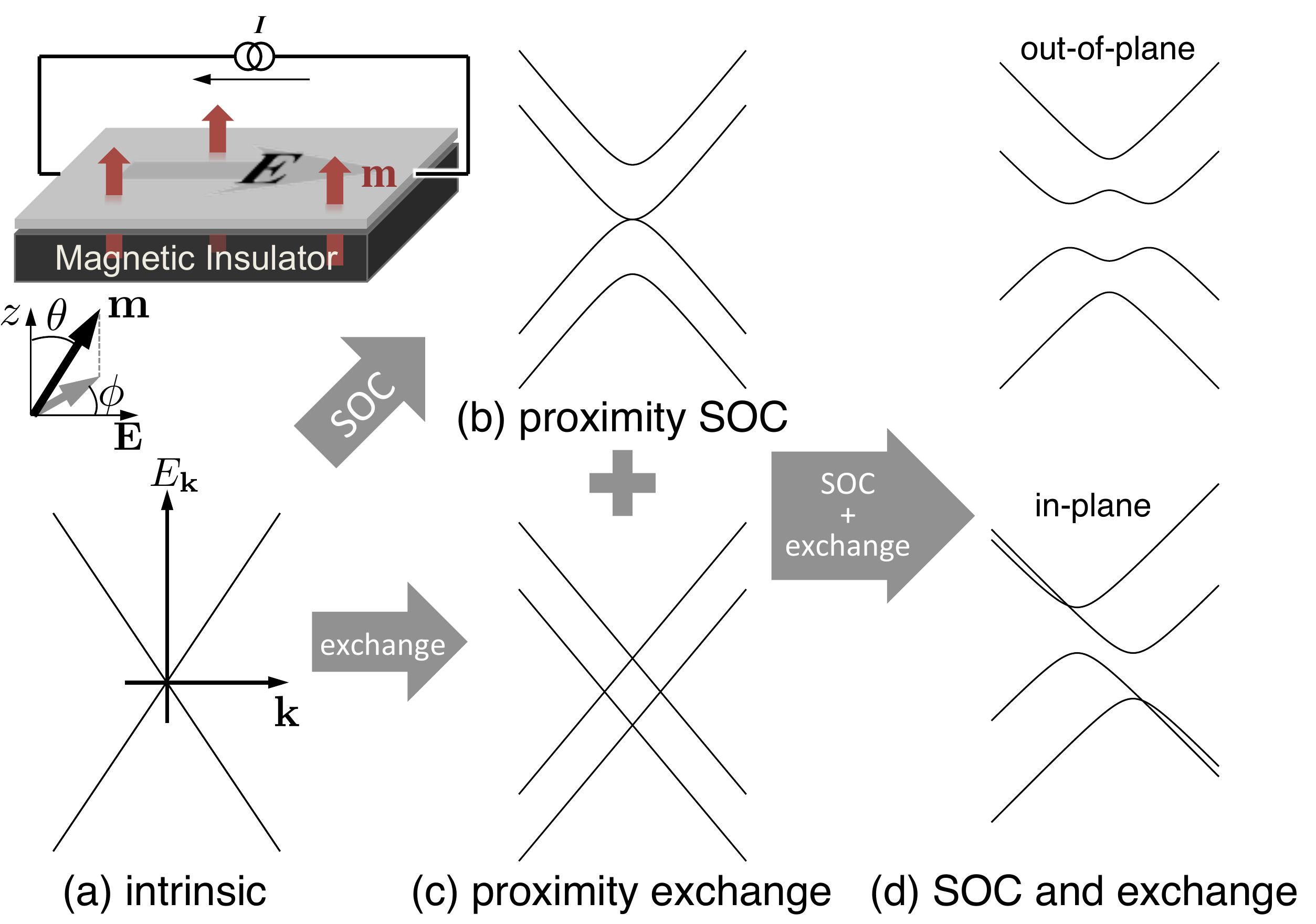}%
\caption{(Color online). Scheme of magnetoanisotropic transport experiment in 
proximity graphene. Polar ($\theta$) and azimuthal ($\phi$) angles define the 
magnetization orientation with respect to the applied electric field.  
(a) Linear energy dispersion of pristine graphene
can be modified by (b) (intrinsic and Bychkov-Rashba) spin-orbit coupling
or (c) exchange field, both leading to spin splitting. (d) The interplay of the two interactions 
makes the bands anisotropic with respect to the magnetization orientation, here shown
as out-of-plane and in-plane. 
\label{fig:01}}
\end{figure}

In this paper 	we solve numerically exactly a realistic Boltzmann transport model, with
long-range charge scatterers, for Dirac electrons in the presence of proximity exchange and spin-orbit couplings. We start with the anisotropic band structure, as in Fig.~\ref{fig:01}, 
and explore its ramifications in transport. Specifically, we introduce and calculate the proximity anisotropic
magnetoresistance, as an analog of the anisotropic lateral magnetoresistance in 
ferromagnetic metal/insulator slabs \cite{Hupfauer:2015hha}, characterizing interfacial spin-orbit fields. We also 
present magnetoanisotropies of the planar Hall effect and inverse spin-galvanic effect.
Finally,  we give parameter maps indicating regions of large transport magnetoanisotropies.

\begin{figure}{}
\includegraphics[width=\columnwidth]{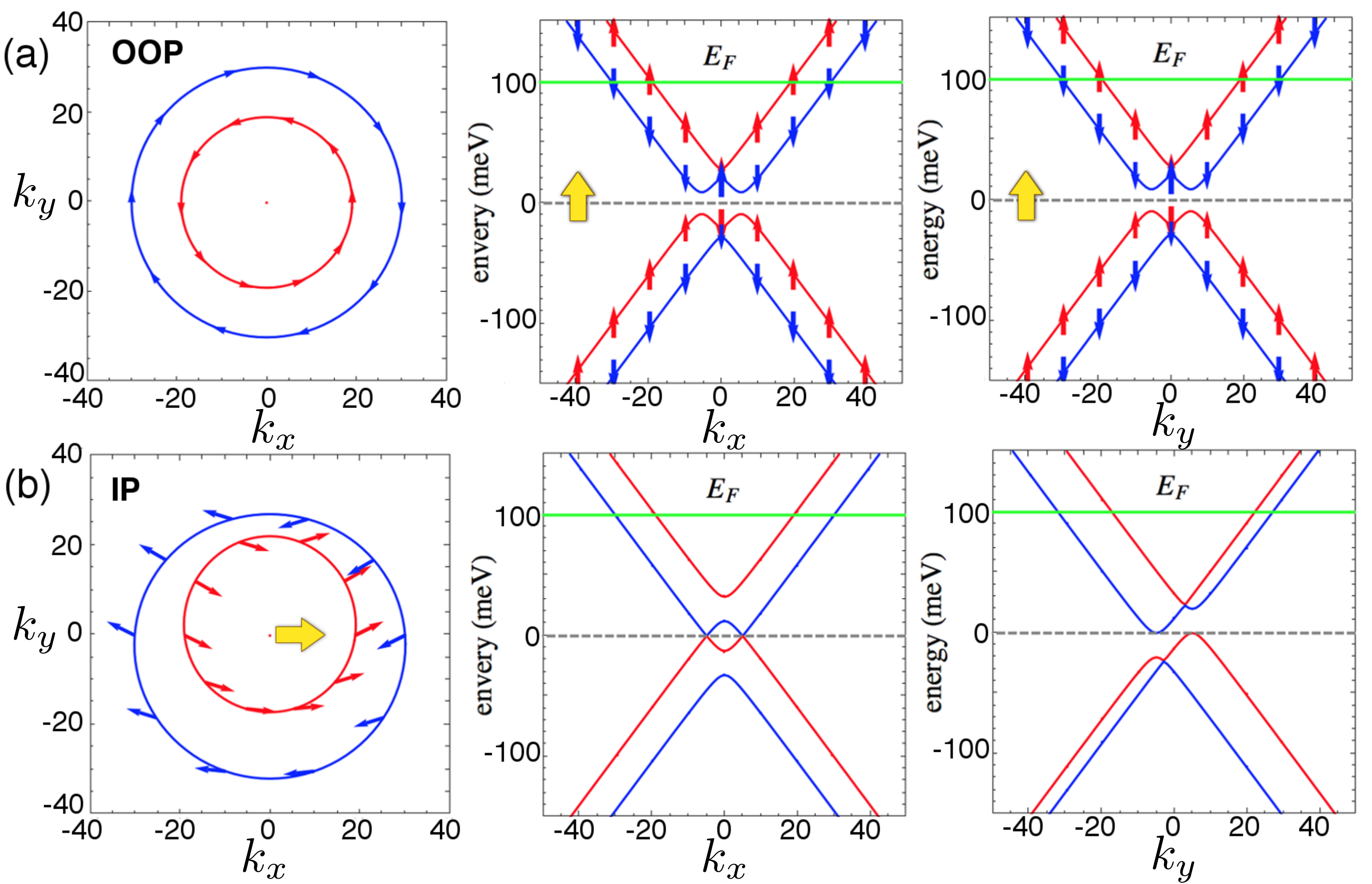}%
\caption{(Color online).
Fermi contours and spin texture (left), and the band structure along $k_x $ (middle) 
and $k_y$ (right) for different directions of the exchange field. 
The direction of the exchange field is indicated by the large arrows; 
small arrows give the spin projections. 
(a) The out-of-plane (OOP) exchange field separates two spin subbands, while
the Bychkov-Rashba field leads to a distinctive spin texture depending 
on the z-projection of the real spin, which interacts with exchange field. 
(b) The in-plane (IP) exchange field splits the bands,
but also deforms the Fermi circles. We used 
$E_F=100\,\text{meV}$, $\lambda_I=0\,\text{meV}$, $\lambda_\text{BR}=10\,\text{meV}$, 
and $\lambda_\text{ex}=25\,\text{meV}$.
\label{fig:02}}
\end{figure}

Dirac electrons in graphene in the presence of proximity exchange and spin-orbit couplings
are described by the minimal Hamiltonian \cite{Han:2014dc},
\begin{equation}\label{eq:01}
H=H_{0}+H_{I}+H_\text{BR}+H_{\text{ex}}.
\end{equation}
Here, pristine graphene Hamiltonian is $H_0=\hbar v_F (\tau_z \sigma_x k_x + \sigma_y k_y)$ with pseudospin (sublattice) Pauli matrices $\pmb{\sigma}$ and $\tau_z=\pm 1$ for 
$K$ and $K'$ points. The Fermi velocity is
$\hbar v_F = (3/2) t a_0 \approx 8.6 \times 10^{7}$ cm/s for $t=2.7$ eV and 
the inter-atomic distance of carbons in graphene $a_0=1.42$ \AA \cite{Peres:2010eb}.
The proximity intrinsic-like spin-orbit coupling is given by the Hamiltonian
$H_I=\lambda_{I} \tau_z \sigma_z s_z$, 
with parameter $\lambda_{I}$ and (true) spin Pauli matrices $\bf{s}$. The intrinsic
coupling opens a gap of $2\lambda_{I}$.  
The Bychkov-Rashba Hamiltonian,  
$H_\text{BR}=\lambda_\text{BR}(\tau_z \sigma_x s_y-\sigma_y s_x)$,
with parameter $\lambda_{BR}$ describes the proximity spin-splitting due to spin-orbit
coupling and lack of space inversion symmetry. Finally, the spin-dependent hybridization 
with the ferromagnet leads to a proximity exchange, 
$H_\text{ex}=\lambda_\text{ex} {\bf m {\cdot} s}$ with parameter $\lambda_{\rm ex}$
and magnetization orientation ${\bf m}$.

There are two important magnetic configurations to consider: out-of-plane and 
in-plane magnetizations, depicted in Fig.~\ref{fig:02} (see also Fig.~\ref{fig:01}). In the out-of-plane case the spin up and spin down bands are spin split, but the band structure (and thus
Fermi contour) remains isotropic. On the other hand, in the in-plane case
the band structure is markedly anisotropic, with the Fermi contours shifted relative to 
each other.

To investigate electrical transport, we solve the linearized Boltzmann equation for the above
model, assuming spatial homogeneity. In the presence of a longitudinal electric field $\bf{E}$, the non-equilibrium distribution function is $f=f_{0} + \delta f$, 
where $f_{0}$ is the equilibrium Fermi-Dirac function. We use the ansatz 
$\delta f=  -e(-\partial f_0/ \partial E) {\bf u}{\cdot}{\bf E}$ and consider long-range
Coulomb scattering, which is the established model for
resistivity in graphene \cite{Adam:2007kc}. The unknown vector ${\bf u}$ is  found by solving the integral equation (obtained from the Boltzmann equation in 
linear order in ${\bf E}$),
\begin{eqnarray}\label{eq:02}
\begin{array}{ll}
{\bf v(k)}
&= 2 \pi n_i (\hbar v_F r_s)^2  \\
&\times\oint_{E_F} \frac{d{\bf k'}}{|\nabla_{\bf k'} E_{\bf k'}|} 
\frac{F({\bf k, k'})}{q^2\varepsilon(q)^2}
\left[
{\bf u (k)}
-
{\bf u (k')}
\right],
\end{array}
\end{eqnarray}
where 
$E_{{\bf k}}$ is the energy corresponding to wave vector ${\bf k}$, ${\bf v (k)}$ is the group velocity, $n_i$ is the concentration of scatterers, 
the effective fine structure constant $r_{s} \approx 0.8$ \cite{Hwang:2007da},
$F({\bf k, k'})=|\Psi({\bf k})^\dagger\Psi({\bf k'})|^2$ 
is the overlap between the incident (${\bf k}$) and scattered (${\bf k}'$) 
states $\Psi$. For example, for  pristine graphene $F({\bf k, k'})=(1+\cos\theta_{\bf k k'})/2$.
For simplicity, spin and pseudospin indices are implicit in the momentum labels 
${\bf k}$. The integral is over the Fermi contour of Fermi energy $E_F$, and 
the transferred momentum is $q = |{\bf k} - {\bf k}'|$. The dielectric function 
$\varepsilon$
is calculated from the random phase approximation \cite{Hwang:2007da, Hwang:2007ky, Hwang:2008gg}.
\begin{eqnarray}
\begin{array}{l}
\varepsilon(q)=\\
\\
{\bigg \{}
\begin{array}{ll}
1{+}\frac{q_{s}}{q} & \textrm{if}\; q{\leq}2k_F\\
1{+}\frac{\pi r_{s}}{2}{+}\frac{q_{s}}{q} {-} \frac{q_s\sqrt{q^2-4k_F^2}}{2q^2}{-}r_s \sin^{-1}(\frac{2k_F}{q})
& \textrm{if}\; q{>}2k_F
\end{array}
\end{array},\quad
\end{eqnarray}
where $q_{s} = 4k_F r_{s}$. The Fermi wave vector $k_F$ is taken from the
pristine graphene case corresponding to a given electron density.  The integral
equation, Eq.~(\ref{eq:02}), is solved numerically exactly \footnote{We discretize the Fermi contour (100 to 500 points usually suffice) and solve the resulting sets of linear  
equations algebraically.}, taking the energy spectrum and eigenstates
of the effective hamiltonian, Eq.~(\ref{eq:01}).

\begin{figure}
\includegraphics[width=0.8\columnwidth]{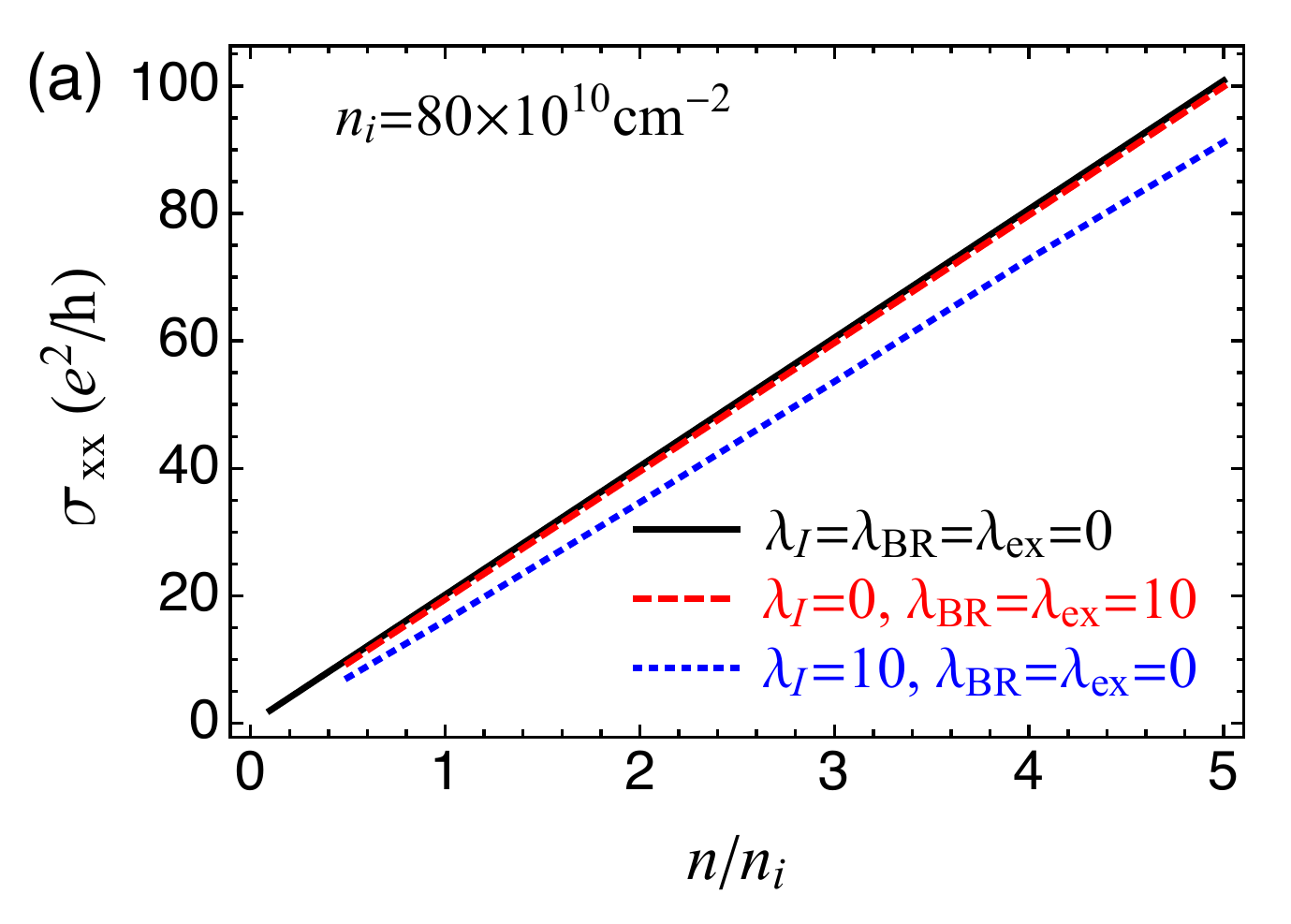}\\
\includegraphics[width=0.8\columnwidth]{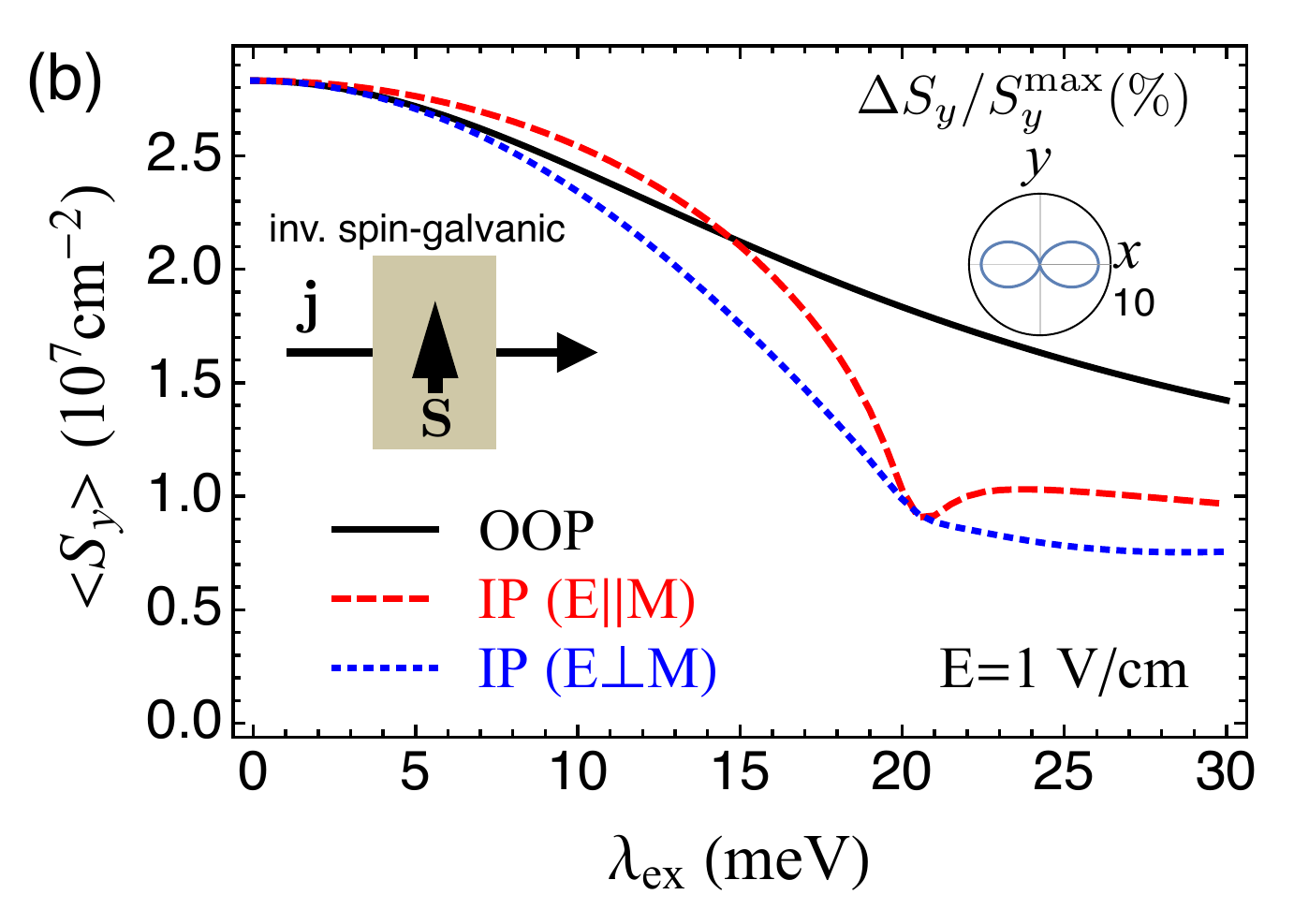}%
\caption{
(a) (Color online). Calculated longitudinal conductivity as a function of carrier density, for pristine and proximity graphene. For proximity graphene we show the conductivity
in the presence of Bychkov-Rashba and exchange coupling only, and in the presence of intrinsic spin-orbit coupling only.
(b) Inverse spin-galvanic effect (scheme in the left inset) in proximity graphene. 
Spin density induced (and normalized) by electric field (which is along x-axis) with respect to the exchange interaction, when the magnetization is out-of-plane (OOP) and in-plane (IP).
In-plane magnetization can be either parallel (along x-ais)
or perpendicular (along y-axis) to the electric field $E$.
In the right inset, the angle dependance of $S_y(\phi)$ is shown in the polar plot
with $\Delta S_y=S_y - S_y^\text{min} $,  
for the electric field $E=1\,\text{V{/}cm}$, and 
for $\lambda_{\textrm{ex}}=10$ meV in percentage with respect to  
$S_y^\text{max}=S_y({\bf E||M})$. The carrier density in this plot is
n = $10^{12}$ cm$^{-2}$ and $\lambda_{I}=0\,\text{meV}$, $\lambda_{\text{BR}}=20\,\text{meV}$.
\label{fig:03}}
\end{figure}

Knowing vectors ${\bf u}$ for the Fermi contour momenta ${\bf k}$, the conductivity tensor is obtained from,
\begin{eqnarray}
\sigma_{ij}= \frac{e^2}{h} \int \frac{d{\bf k}}{2\pi} \hbar v_{i} u_{j} \delta(E_{\bf k}-E_F).
\label{eq:04}
\end{eqnarray}
We plot the calculated longitudinal conductivity of graphene as a function of 
carrier density $n$,  
with and without proximity effects, in Fig.~\ref{fig:03}(a). We use  $n_i= 80\times 10^{10}\,\text{cm}^{-2}$ as  a representative density of long-range scatterers. 
The carrier density, unlike in pristine graphene, depends not only on  
the Fermi level but also on the strength of the proximity interactions, 
$ \lambda_I$ and $\lambda_{\textrm{ex}}$,
\begin{eqnarray}
n(E_F)=2 \times
\frac{1}{2\pi}
\left[
E_F^2+2E_F \lambda_I +\lambda_{\textrm{ex}}^2
\right]/(\hbar v_F).
\end{eqnarray}
The factor 2  takes into account the valley degeneracy.
The carrier density is independent of $\lambda_\text{BR}$
and the direction of the magnetization. In all the plots we fix the carrier
density, instead of the Fermi level. 
The conductivity for three different combinations of parameters is 
shown in Fig.~\ref{fig:03}(a).  The linear dependence on $n$ is well reproduced. 
While $\lambda_\text{BR}$ and $\lambda_\text{ex}$ bring about 
relatively insignificant changes ($\lesssim$ 2.5 \%), 
the presence of  $\lambda_{I}{=}10\,\text{meV}$ lowers the conductivity
by about ${\sim}10$ \% at a fixed carrier density. Thus, in terms of modifying the magnitude of the conductivity, the proximity effects  (unless not inducing additional scattering, which would need to be investigated case-by-case) are rather weak, 
being more pronounced with the inclusion of the intrinsic spin-orbit coupling, 
than with the Bychkov-Rashba and exchange effects. However, as we will see 
shortly, the anisotropic effects are quite pronounced.

\begin{figure}[t]
\includegraphics[width=0.8\columnwidth]{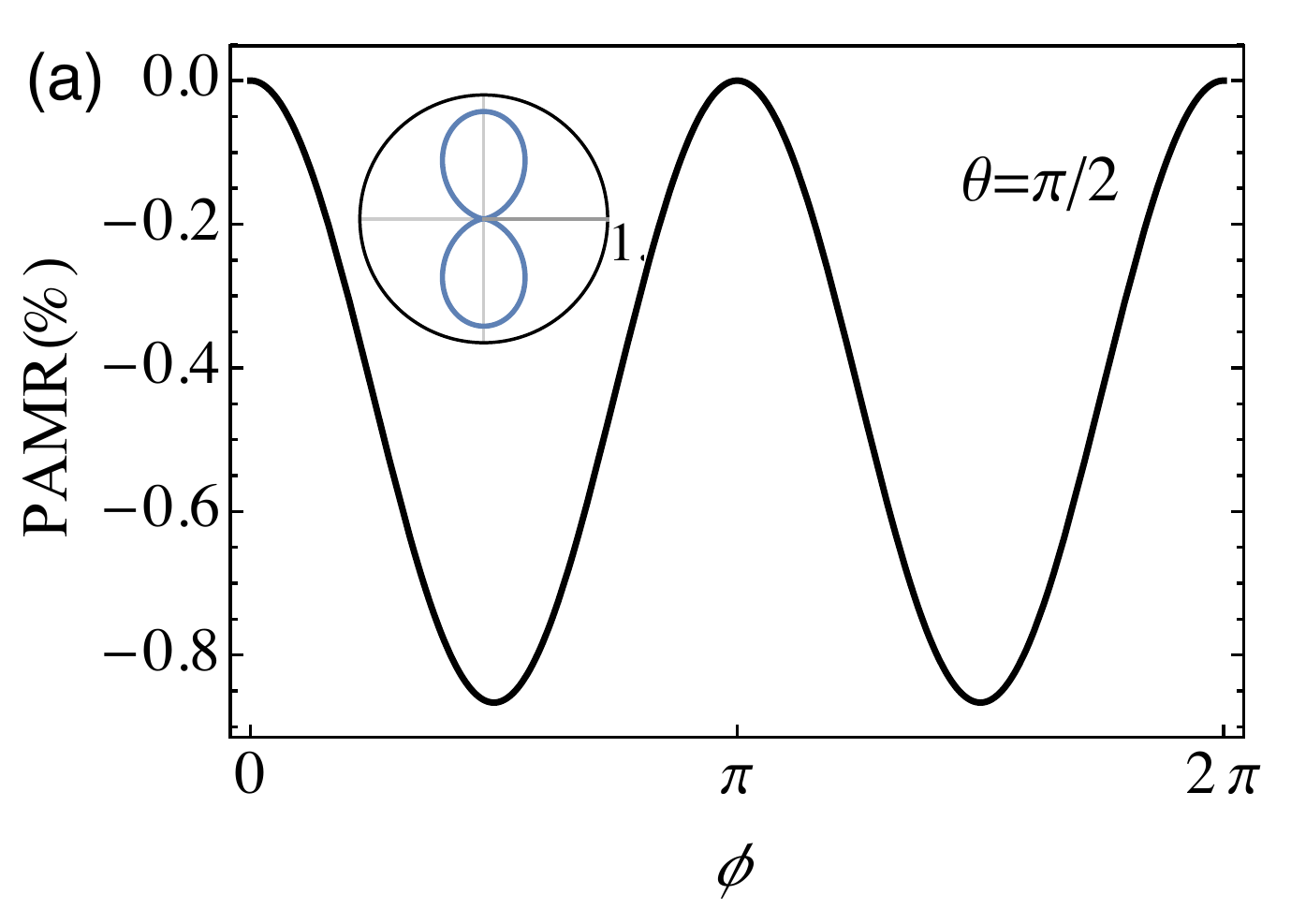}\\
\includegraphics[width=0.8\columnwidth]{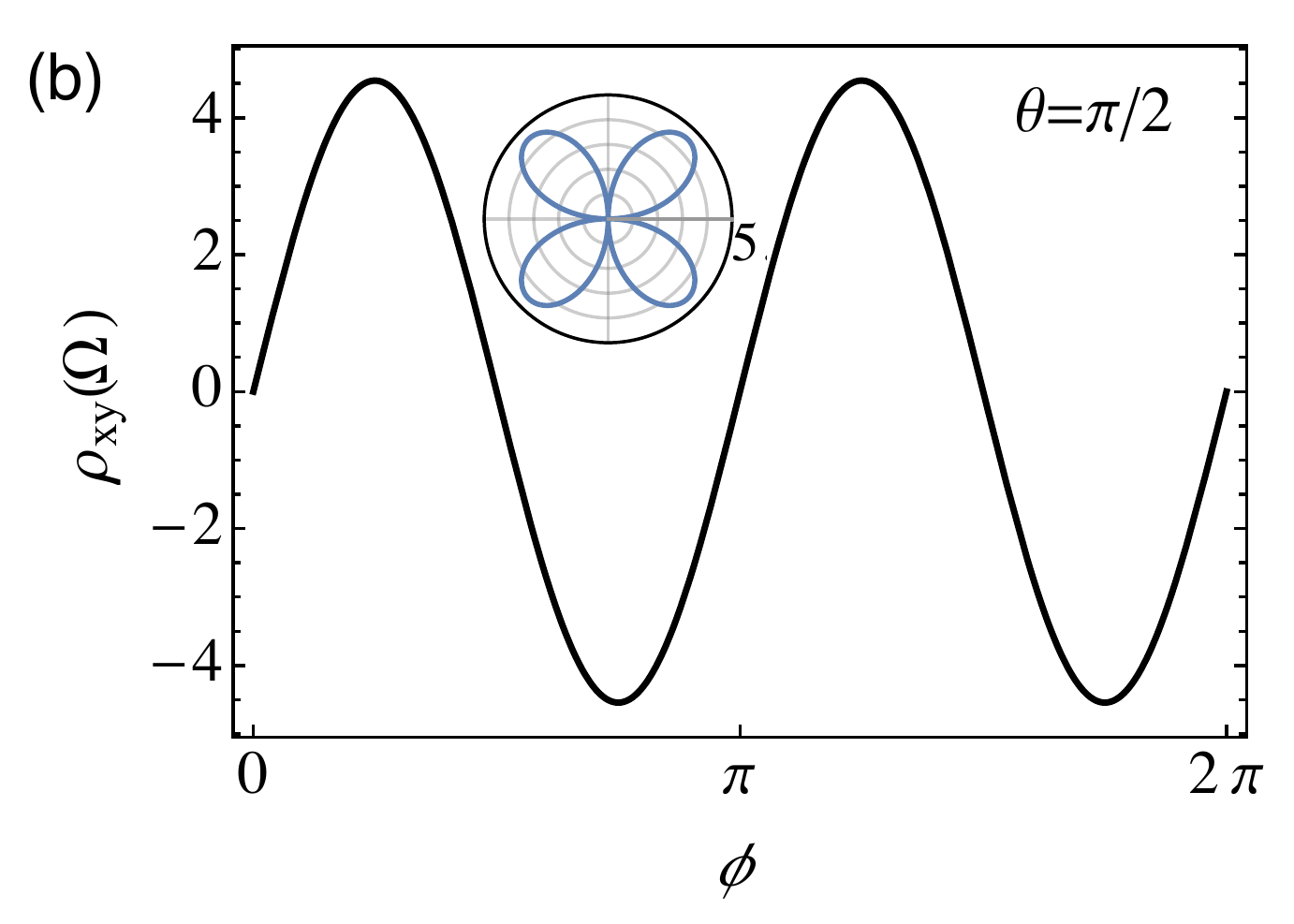}%
\caption{(Color online).
(a) Calculated proximity induced anisotropic magnetoresistance (PAMR) and (b) planar Hall resistivity are shown
when the exchange field is in-plane ($\theta=\pi/2$) for $\lambda_\text{BR}=20\,\text{meV}$.
PAMR quantifies the  longitudinal magnetoresistance
as a function of magnetization orientation $\phi$. The interplay of spin-orbit coupling and exchange field leads to a net 
anisotropic resistivity with $C_{2v}$ symmetry while the off diagonal elements
of the resistivity tensor are non-zero.
Other parameters are $n=10^{12}\,\text{cm}^{-2}$,
$\lambda_I=0\,\text{meV},\,\lambda_\text{ex}=10\,\text{meV}$.
The insets are  polar plot representations.
\label{fig:04}}
\end{figure}

\begin{figure*}[t]
\includegraphics[width=\columnwidth]{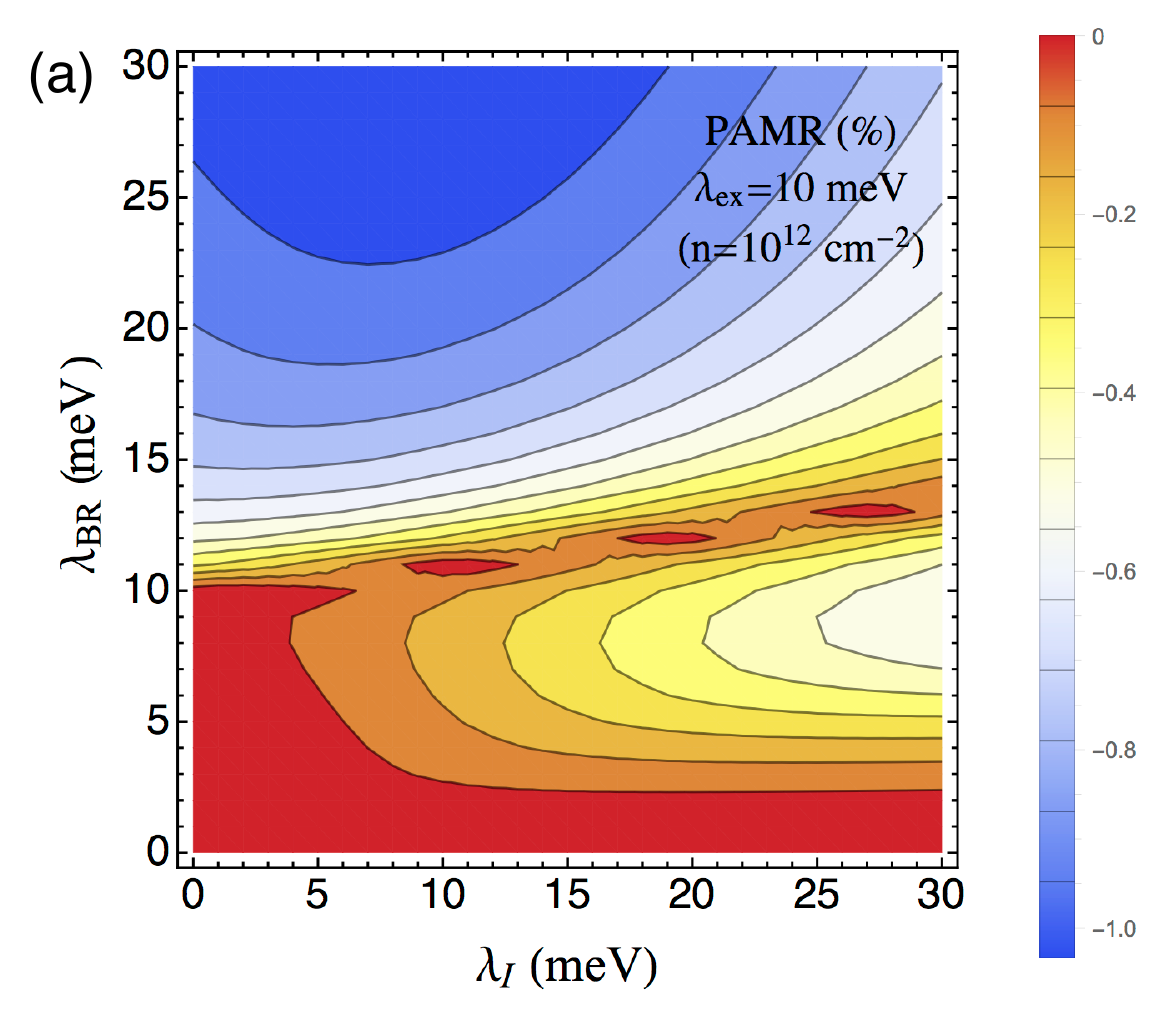}%
\includegraphics[width=\columnwidth]{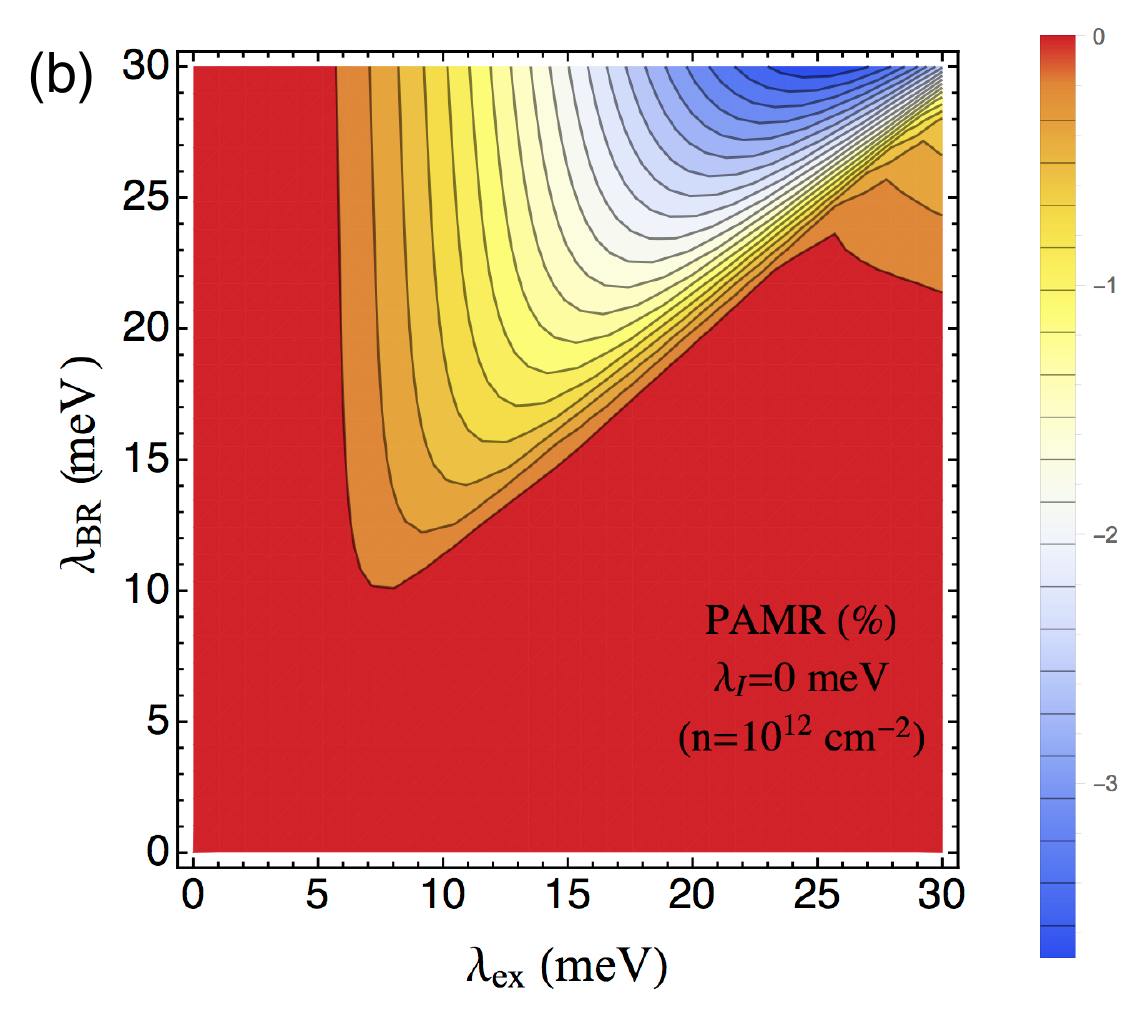}%
\caption{(Color online). Parameter maps of PAMR($\theta=\pi/2$, $\phi=\pi/2$).
(a) PAMR as a function of $\lambda_\textrm{BR}$
and $\lambda_{I}$, for a fixed $\lambda_\textrm{ex}=10$ meV.
(b) PAMR as a function of $\lambda_\textrm{BR}$
and $\lambda_\textrm{ex}$, for $\lambda_{I} = 0$.
Note that PAMR changes sign around the $\lambda_\textrm{BR}=\lambda_\textrm{ex}$ line. In both maps the carrier density is $10^{12}$ cm$^{-2} $.
\label{fig:05}}
\end{figure*}

When current flows in the presence of a Bychkov-Rashba field, a spin density transverse
to the current appears as a demonstration of the inverse spin-galvanic effect \cite{Ganichev:2002eu, Ganichev:2006th, Ganichev:2014ej} . The shift of the spin subbands due to the 
electric field, combined with the spin texture due to the Bychkov-Rashba spin-orbit interaction, leads to a spin polarization. This is also expected to happen 
in graphene \cite{Dyrdai:2013ey, Dyrdai:2014be}, along with the spin-galvanic
effect \cite{Shen:2014cb, Sanchez:2013kb}.  
The non-equilibrium spin density caused by the electric field can be
calculated as,
\begin{eqnarray}
\delta\langle{\bf S}\rangle= \int \frac{d{\bf k}}{(2\pi)^2} \delta f({\bf k}) {\bf s({\bf k})},\label{eq:06}
\end{eqnarray}
where ${\bf s(k)}$ is the spin (represented by Pauli matrices $\bf s$) 
expectation value of the state ${\bf k}$. For our proximity
model in the presence of long-range Coulomb scatterers, the calculated inverse
spin-galvanic effect is shown 
in Fig.~\ref{fig:03}(b) as a function of exchange coupling. 
With increasing magnetization the induced transverse spin is
reduced, as the exchange coupling aligns the spins and deforms 
the rotational spin texture of the Bychkov-Rashba field. The spin densities
can be giant. In fields of 1 V/$\mu$m, which are still achievable in graphene, 
the spin density could reach $10^{11}$ cm$^{-2}$, corresponding to about 10\% of
spin polarization.  The largest induced spin accumulation is in the out-of-plane
configuration for large exchange. The magnetoanisotropy of the inverse spin-galvanic
effect can be very large, as seen in  Fig.~\ref{fig:03}(b). The presence of the
current-induced spin accumulation, as well as its magnetoanisotropy, could be detected in the same proximity structure, by measuring 
the transverse voltage, as in nonlocal spin injection \cite{Zutic:2004fo}.

To quantify the transport magnetoanisotropy, we introduce {\it 
proximity induced anisotropic magnetoresistance (PAMR)}, as a ratio
of the resistivities $R$ (or conductivities $\sigma$) for a given magnetization orientation
$(\theta, \phi)$ (see Fig.~\ref{fig:01}),
\begin{equation}
\begin{array}{ll}
\text{PAMR}_{[{\bf E}]}&=\dfrac{R(\theta,\phi)-R(\theta,0)}{R(\theta,0)} \\
&= \dfrac{\sigma_{xx}(\theta,0)- \sigma_{xx}(\theta,\phi)}{\sigma_{xx}(\theta,\phi)},
\end{array}
\end{equation}
analogously to the tunneling anisotropic magnetoresistance effect 
\cite{MatosAbiague:2009gp, MatosAbiague:2009el, Brey:2004dr}.
PAMR refers to the changes in the longitudinal magnetoresistance 
as the magnetization direction varies with respect to the direction of the external electric field that
drives the current. 
When the magnetization is out-of-plane ($\theta=0$),
the broken time reversal symmetry and strong spin-orbit coupling 
can lead to the novel quantum anomalous Hall effect \cite{Qiao:2014fw}, 
or crystalline magnetoanisotropy \cite{Hupfauer:2015hha}.
However, here we focus on the regime in which PAMR is most pronounced ($\theta=\pi/2$). As shown in 
Fig.~\ref{fig:04}(a), PAMR exhibits a $C_{2v}$ symmetry due to the 
interplay between the Bychkov-Rashba and exchange interactions.
The expected magnitudes of PAMR are about 1\%, similar to what
is observed in ferromagnetic metals \cite{Hupfauer:2015hha}. 
The anisotropic resistivity tensor has non-zero off-diagonal elements due to the
presence of  exchange and spin-orbit couplings. This leads to  
the planar Hall effect, shown in Fig.~\ref{fig:04}(b).
The magnitude of the planar Hall effect could reach up to 4 $\Omega$
which greater than the typical 
values studied in metallic ferromagnetic systems \cite{Tang:2003ku}.

What values can PAMR reach for a reasonable range of proximity parameters?  
Figure~\ref{fig:05} shows two parameter maps, one with the Bychkov-Rashba and 
intrinsic, the other with the Bychkov-Rashba and exchange couplings. We see
two distinct features. (i) First, in Fig.~\ref{fig:05}(a) a horizontal line around $\lambda_\text{BR}\sim \lambda_{\text{ex}} \approx 10 $ meV separates two regions. For $\lambda_\text{BR}\lesssim10$ meV,
increasing $\lambda_{I}$ increases PAMR. For $\lambda_\text{BR}\gtrsim 10$ meV, PAMR
initially increases with increasing $\lambda_{I}$, reaching a maximum of about 
1 \% around $\lambda_{I}\sim 7$ meV, beyond which PAMR decreases.
(ii) Second, in Fig.~\ref{fig:05}(b) the line $\lambda_\text{BR}=\lambda_\text{ex}$ 
marks a sharp crossover between weak and strong PAMR. However, this crossover
is not uniform. PAMR is largest for large values of both $\lambda_\text{ex}$
and $\lambda_\text{BR}$ slightly greater than $\lambda_\text{ex}$. The reason why this
region gives the largest PAMR (more than 1\%) is that in this parameter 
range there is a band crossing 
between the strongly spin-orbit coupled subbands.

In conclusion, we used a realistic transport model to predict magnetotransport 
anisotropies in graphene with proximity exchange and spin-orbit couplings. 
We predict marked anisotropies in the magnetoresistance, with similar values as
reached in ferromagnetic metal junctions and slabs. The calculated
PAMR depends
strongly on the spin-orbit coupling and exchange parameters.  We also calculated
the magnetoanisotropies of the planar Hall and inverse spin-galvanic effects. All these
magnetoanisotropies should  be a sensitive tool to probe proximity effects in graphene.

\begin{acknowledgments}
We thank Denis Kochan, Jonathan Eroms, and Cosimo Gorini for useful discussions. 
This work was supported by the DFG SFB 689 and the European Union Seventh Framework Programme under Grant Agreement No. 604391 Graphene Flagship.
\end{acknowledgments}

\bibliography{PAMR}

\end{document}